\documentclass[twocolumn,aps,superscriptaddress]{revtex4}
\usepackage{amssymb}
\usepackage{amsfonts}
\usepackage{amsmath}
\usepackage{amsxtra}
\usepackage{amscd}
\usepackage{bm}
\usepackage{amsthm}
\usepackage{graphicx}
\usepackage{epsf}
\usepackage{bbold}
\usepackage{xcolor}
\usepackage{mathalfa}
\usepackage{makecell} 
\usepackage{multirow}
\usepackage{hhline}
\usepackage{tikz}
\usepackage{physics}
\usepackage{eucal}
\usepackage{stackengine}

\newcommand{\eq}[1]{Eq.(\ref{#1})}

\newcommand{\beq}{\begin{equation}}
\newcommand{\eeq}{\end{equation}}
\newcommand{\bmul}{\begin{multline}}
\newcommand{\emul}{{\end{multline}}}
\newcommand\beqa{\begin{eqnarray}}
\newcommand\eeqa{\end{eqnarray}}
\newcommand\bea{\begin{array}}
\newcommand\eea{\end{array}}
\newcommand\ba{\begin{array}}
\newcommand\ea{\end{array}}
\newcommand{\nn}{\nonumber}
\newcommand{\eV}{{\,\rm eV}}
\newcommand{\meV}{{\, \rm meV}}
\newcommand{\T}{{\, \rm T}}

\newcommand{\neqa}{\nonumber\end{eqnarray}}

\newcommand{\nm}{{\, \rm nm}}

\begin{document}

\title{Electronic properties of stacking faults in Bernal graphite}

\author{Patrick Johansen Sarsfield}
\affiliation{National Graphene Institute, University of Manchester, Manchester M13 9PL, United Kingdom}
\affiliation{Department of Physics \& Astronomy, University of Manchester, Manchester M13 9PL, United Kingdom}

\author{Sergey Slizovskiy}
\affiliation{National Graphene Institute, University of Manchester, Manchester M13 9PL, United Kingdom}
\affiliation{Department of Physics \& Astronomy, University of Manchester, Manchester M13 9PL, United Kingdom}

\author{Mikito Koshino}
\affiliation{Department of Physics, Osaka University, Toyonaka, Osaka 560-0043, Japan}

\author{Vladimir Fal'ko}
\affiliation{National Graphene Institute, University of Manchester, Manchester M13 9PL, United Kingdom}
\affiliation{Department of Physics \& Astronomy, University of Manchester, Manchester M13 9PL, United Kingdom}

\email{Vladimir.Falko@manchester.ac.uk}


\begin{abstract}
Using the tight-binding model of graphite, incorporating all Slonczewski-Weiss-McClure parameters, we compute the spectrum of two-dimensional states of electrons bound to a stacking fault in Bernal graphite. We  find that those bands retain characteristic features of the low-energy bands of a rhombohedral graphene trilayer, which actually represents the lattice structure  the fault.
Based on the self-consistent analysis of  charge and potential distribution across the fault layers,  we determine the shape of the Fermi contour for the 2D band, which has the form of three pockets with a  hole-like conic dispersion and Dirac points above the Fermi level. The computed frequency of  Shubnikov-de Haas oscillations and the cyclotron mass of the fault-bound
charge carriers (at the Fermi level) are sufficiently different from the corresponding bulk values in graphite, making  such stacking faults identifiable by quantum transport and cyclotron resonance measurements.
\end{abstract}

\maketitle

\section{Introduction}
Despite almost a century-long history of research in physical  properties of graphite \cite{McClure1957,McClure1960,DresselhausReview}, this material continues to bring up unexpected observations \cite{MLG2004, BLG2006, Artem2019}. Owing to a weak van der Waals intelayer coupling, graphite exists in several  polytypes, of which Bernal (AB) and rhombohedral (ABC) are the best studied in terms of their electronic properties \cite{DresselhausReview,Artemlatychevskaia2018,Artem2019,ArtemABC2020}. One half of carbons in Bernal graphite are arranged into chains aligned with the vertical direction ("double-dimers"), whereas the other half appears to be in between empty centers of hexagons in graphene layers above and underneath.  This determines the electronic properties of bulk graphite to be semimetallic  with a Fermi surface that contains electron and hole pockets, as confirmed by  Shubnikov-de Haas (SdH) oscillations \cite{Soule1964,DresselhausReview,SchneiderPotemski2009,schneider10,SchneiderPotemski2009,schneider2012using} and cyclotron resonance  (CR) measurements \cite{McClure1957, Galt56,Nozieres57,Inoue62,Ushio72,Suematsu72,DresselhausReview,Potemski2012}. In contrast, all carbons in rhombohedral graphite lattice are arranged into pair which appear on the top of each other in the consecutive layers (so-called "dimers"), which makes thin films of rhombohedral graphite insulating in their bulk, due to interlayer hybridization of carbon's $P^z$ orbitals. This films also feature  surface states, which are  topologically protected in the spirit of Su-Schrieffer-Heeger model \cite{SSH} and have already been observed experimentally \cite{yankowitz2014electric, Yankowitz2019, ArtemABC2020} demonstrating  2D semimetallic properties with a clearly manifested electron-hole asymmetry \cite{slizovskiy_films_2019}.

Statistically, each of those bulk polytypes of graphite may contain sporadic stacking faults, modifying electronic properties of bulk material or a thin film. Here we present a model for electronic properties of an isolated  stacking fault in Bernal graphite. Structurally, such a fault can be viewed as an ABC trilayer surrounded by Bernal bulk crystal, see Fig.\ref{fig:1}a. While in Fig. \ref{fig:1}a the fault looks like an offset of carbon chains in the upper and lower parts of semi-infinite Bernal crystals, however, as noted in Ref.\cite{Muten_2021}, such structural fault can not be generated by a mutual shift of two Bernal half-spaces. Moreover,  due to the ABC-structure of the fault, it hosts  localized two-dimensional (2D)  states, inherited from the surface states of the rhombohedral trilayer, noted in several recent publications \cite{Arovas2008,Muten_2021,Taut2013,Taut2014,Taut2016,ABC_fault2023}.
In this work we study the electronic properties of such a  stacking fault in order to describe in detail the dispersion characteristics of the 2D bands that it hosts, in particular, to predict experimentally measurable characteristics such as the period of Shubnikov-de Haas oscillations and the cyclotron mass of 2D carriers at the Fermi level.
In contrast to the earlier theories of fault-bound 2D bands, here, we account for the full set of Sloczweski-Weiss-McClure (SWMcC) parameters for graphite \cite{Slonczewsi1958,McClure1957,McClure1960} implemented in the hybrid $\bm{ k\cdot p}$ - tight binding theory, supplemented by the self-consistent Hartree computation of on-layer energies.  
Taking all SWMcC parameters into account (in particular, the next-neighbor/layer hoppings and coordination-dependent on-carbon potentials) is crucially important, as those lift an artificial degeneracy of band edges (similarly to what has been found in monolithic ABC films \cite{Koshino2009trigonal,slizovskiy_films_2019}) which haunt the minimal model with only the closest neighbor hopping \cite{Muten_2021,Arovas2008}. The full set of SWMcC parameters in the model  is also needed for correctly describing trigonal warping of the electronic band structure of bulk, surface and 2D fault states. As a result, for the fault-localized 2D bands, we identify a set of three distinct Fermi pockets characterized by SdH frequency $\nu_{SdH}\approx 1\T$  (distinctively different from the values  $\nu_{SdH}^{e} \approx 6.3\T$ and $\nu_{SdH}^{h}\approx 4.6\T$ reported for bulk Bernal graphite \cite{Soule1964,DresselhausReview,SchneiderPotemski2009,schneider10,SchneiderPotemski2009,schneider2012using})  and cyclotron mass of $m_{CR}\approx 0.04 m_0$  (to be compared to $m_{CR}\approx 0.06 m_0$ related to CR of bulk electrons \cite{McClure1957, Galt56,Nozieres57,Inoue62,Ushio72,Suematsu72,DresselhausReview,Potemski2012}, where $m_0$ is the free electron mass). We also identify the differences between inter-Landau-level selection rules for cyclotron transitions in the 2D bands and in the bulk, and analyze the dependence of the fault characteristics on the values of fault-related tight-binding parameters. 
   
\section{Results and discussion}
\subsection{Tight-binding model for a fault in Bernal graphite.\label{sec:2A}}

\begin{figure*}[]
\centering
  { \includegraphics[width=0.8\textwidth]{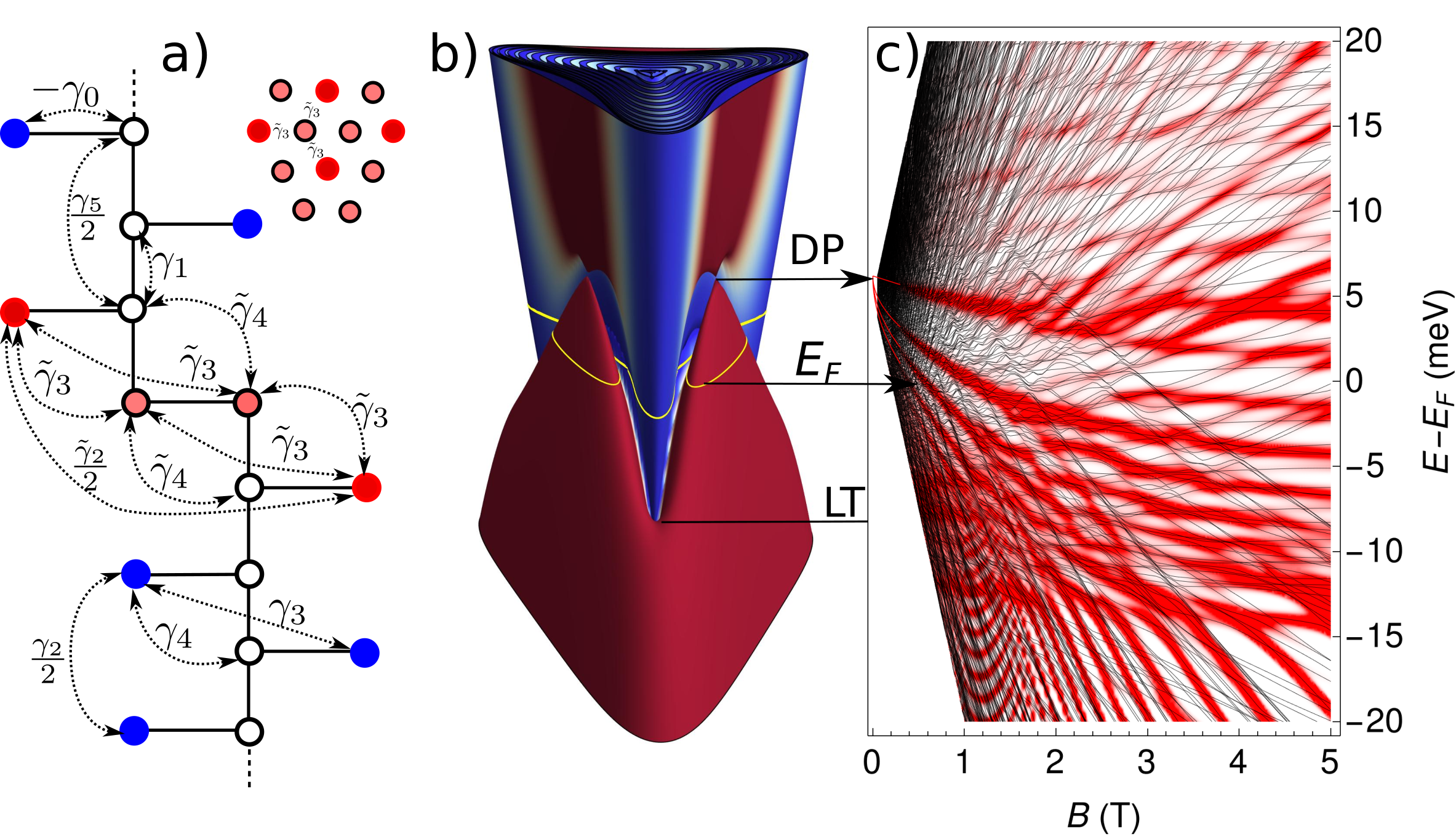}}
\caption{\label{fig:1}
a) Schematic of the ABC stacking fault embedded in Bernal graphite, with labeled hopping parameters. White-colored carbon orbitals correspond to high-energy dimer sites. Blue denotes non-dimer Bernal sites, and red indicates states localized on a stacking fault.  
b) Low-energy bands of a 30+1+30 layer system with red and blue color indicating states localized on 3 ABC stacked layers of a stacking fault and the bulk graphite states respectively. Yellow contour is the Fermi level ($E=0$) for an undoped system.
The fault-bound 2D band is identified by the red colour and it features the Dirac points (DP) in three conic branches which merge together at energies below the Lifshitz transition (LT).  
c) Corresponding Landau levels (black thin lines) superimposed on a density of states  projected onto 3 ABC stacked layers of a stacking fault (in red). Horizontal arrows indicate the states in the band structure which develop into their corresponding  LLs in finite $B$ field.  
}
\end{figure*}

The structure of an ABC stacking fault is sketched in Fig. \ref{fig:1}a, where we also label all the involved SWMcC tight-binding model 
parameters. In this sketch we use the following color-coding: The white-colored sites correspond to carbon dimers hybridized by the strongest interlayer coupling $\gamma_1$.  Non-dimer sites of Bernal graphite are identified using blue color. The sites marked by red are those which host the localized states at the fault with energies close to the Fermi level of graphite. 

The primary choice of the values of SWMcC parameters is based on the values determined for bulk graphite in Ref. \cite{Potemski2012}:
 $\gamma_0 = 3.16\eV$, $\gamma_1=375 \meV$, $\gamma_2=-20 \meV$, $\gamma_3 = 315 \meV $, $\tilde \gamma_{3} = 315 \meV$, $\gamma_4 = 44 \meV$, $\tilde\gamma_4 = 315 \meV$, $\gamma_5=38 \meV$
\cite{Gamma3Sign,yin_dimensional_2019, Ge2021Control,Potemski2012} (with graphene lattice constant $a=0.246$nm).
This choice was made because this parameter set describes well CR and SdH experiments (for details see Fig.\ref{suppfig:parameters}).
While in this primary set we do not distinguish between bulk hopping parameters and those within the fault layers, however, we later account for possible differences between "non-dimer" to "non-dimer" hopping $\gamma_{3}$ {\it vs} "non-dimer" to "half-dimer" hopping $\tilde\gamma_{3}$.  These parameters are implemented in the $k\cdot p$-theory -- tight-binding Hamiltonian \eq{Hamiltonian} described in Supplementary Information. We also use the inset in Fig.\ref{fig:1}a to show the top view of the set of red-painted cites of the ABC trilayer at the fault to illustrate trigonal symmetry of the structure which implies that $\tilde \gamma_{3}$ parameters are the same  for all "non-dimer" to "half-dimer" hoppings. In addition, we allow for a variation of fault-related hopping $\tilde\gamma_{2,4}$ from the bulk values
($\gamma_{2,4}$), due to the difference of their nearest-neighbour environment.  In addition we account for the onsite energy shift, $\Delta'=25 \meV$ produced for each carbon $P^z$ orbital by another carbon above or below: this will produce an energy shift $2\Delta'$ for all dimer sites (white circles) and $\Delta'$ for all sites in the middle plane of the fault (light red).

As localization of electron states at the fault is associated with charge redistribution between nearby layers and the bulk, it is necessary  to account for the resulting variation of on-layer energies self-consistently. Here we implement it using Hartree approximation. A representative example of self-consistently calculated density and on-layer potential distribution is shown in Fig. \ref{fig:Hartree} in the supplementary material for the primary choice of tight-binding model parameters. 

To realize numerical analysis of electronic states of a single fault 
we numerically compute the spectra of graphitic film with 61 layers where the fault resides exactly on the middle layer of the film. These numerical calculations produce a combination of bulk bands, surface states \cite{yin_dimensional_2019,Mullan2023} and fault-localized bands. We also repeat the calculations for the same system in magnetic field and analyze its Landau level (LL) spectrum. In particular, projecting the LL states onto three layers that constitute the fault structure. For graphic representation of the data, we applied broadening $\Gamma=0.2 \meV$ of all LLs which enables us to plot the density of states (DoS) in this system, in particular, local DoS on the fault (projected onto the 3 middle layers of the film). In parallel, we compute the 3D band structure (both at $B=0$ and finite magnetic field) of a 3D "crystal" constructed of a periodic sequence of faults and anti-faults separated from each other by 23 layers of Bernal graphite. In this case, for each in-plane momentum ($k_x,k_y$), the extended bulk states produce broad "1D" bands of $k_z$ dispersion (with negligible gaps), whereas the states localized on the fault can be identified as exponentially narrow (non-dispersive in $k_z$) bands. The latter manifest themselves as peaks in the overall density of states, so that they give $E(k_x,k_y)$ dispersion of fault bound states even without the need to project their wave functions onto the fault layers.
The agreeable comparison of the results of these two methods (with the same broadening of all states applied in both cases) gives us a confidence in the identified 2D dispersion and LL spectrum of the fault-related states.

\subsection{Fault-related electronic dispersion and Landau levels.}

 In Figure \ref{fig:1}b  we display the computed dispersions, $E(k_x,k_y)$, of all 2D sub-bands (representative of size-quantization of electron states in the film along the K-H line  in the 3D Brillouin zone of graphite) in a  film with 61 layers and the stacking fault exactly in its middle. In this plot, we use color-coding to distinguish between bulk (blue) and fault-located (red) sub-bands, which were identified by projecting their wave functions onto the three middle layers of the film. The fault-related states feature three cone-like dispersions (anisotropic due to trigonal warping and with a pronounced electron-hole asymmetry), each ending with a Dirac-like singularity (DP) at the energy $E-E_F\approx6$ meV above the self-consistently calculated Fermi level (the latter is dominated by the bulk states). Such a feature is reminiscent of bands characteristic for an ABC trilayer \cite{Koshino2009trigonal} (illustrated in supplementary Fig. \ref{fig:sup_trilayer}).  These Dirac-like features are characterized by Berry phases $\pi$, which we confirmed by calculating $\phi = -\oint \Im \langle \Psi|\nabla_{\bm p} |\Psi\rangle \cdot d{\bm p}  $ along each of the three Fermi-lines using the wave-function computed for those 2D bands. 
 The apparent  electron-hole asymmetry of the 2D spectrum is the result of the upper branches of the ABC trilayer spectrum blending into the bulk states at the energies  above $6$ meV. The result in Fig. \ref{fig:1}b demonstrates that the  hole-type 2D bands are well isolated from the bulk bands across the energy range $\sim \pm5 \meV$ near the Fermi level, with the fault based bands  traceable down to much lower energies where the individual Dirac cones merge into one single dispersion branch, upon experiencing a Lifshits transition (LT) at $E-E_F\approx-14$ meV. This is inherited from the ABC-trilayer spectrum too, together with Berry phase $3\pi$ that we computed for the constant-energy cut through the spectrum below the LT. Such a behavior was reproduced by the band analysis in a 3D "crystal" constructed of a periodic sequence of faults and anti-faults separated from each other by 23 layers of Bernal graphite. To mention, in the opposite corner of graphite's Brillouin zone (line K'-H') this anisotropy of the dispersion will be inverted as prescribed by time inversion symmetry. 
 
\begin{figure}
\begin{center}
\includegraphics[width=1\columnwidth]{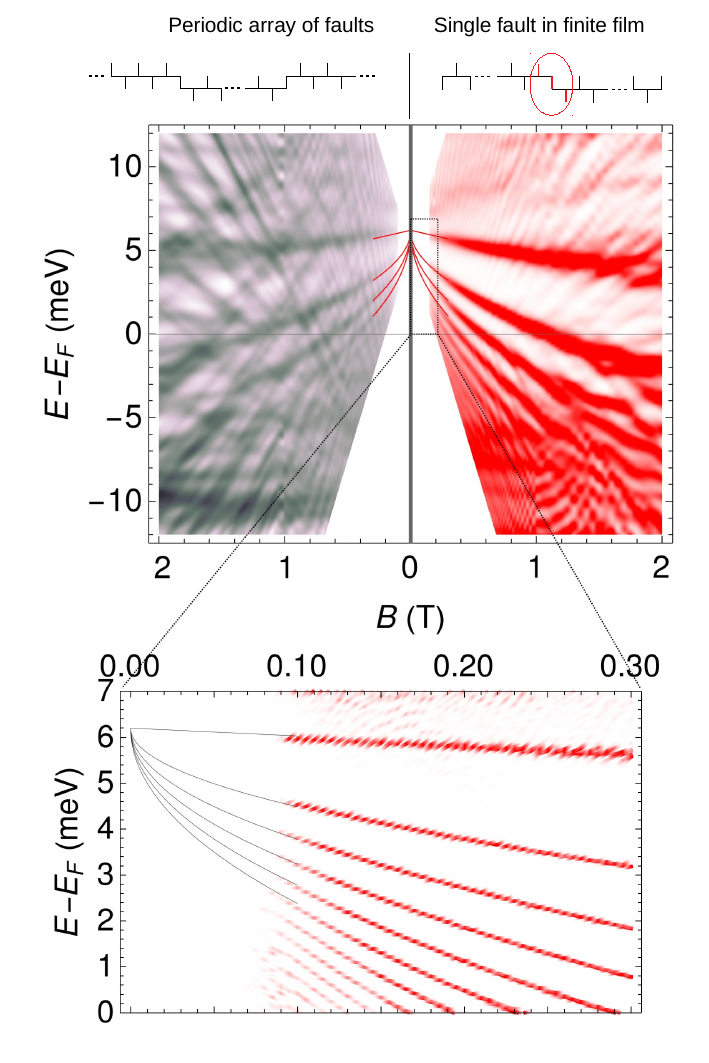}
\caption{(left) The total density of states for an infinite periodic array of stacking faults, each separated by 23 Bernal-stacked layers in a finite magnetic field. (Right) Density of states projected onto the three ABC-stacked layers, calculated for a finite 30+1+30 film with a single stacking fault. Lower image is a zoom at low magnetic fields.   
\label{fig:2}}
\end{center}
\end{figure}

This band structure is mirrored by the magnetic field dependence of  the local DoS on the stacking fault shown in Fig. \ref{fig:1}c with red color, where we observe a characteristic $-\sqrt{n B}$ fan of triply degenerate LLs originating from $E-E_F \approx 6 \meV$ corresponding to the three Dirac points. In addition to the layer projection, we applied broadening $\Gamma=0.2 \meV$ to all computed LLs. The zoomed-in panel in Fig. \ref{fig:2} shows that, at small magnetic fields, we can trace the convergence of these graphene-like 2D LLs towards Dirac points identified in the $B=0$ spectrum in Fig. \ref{fig:1}. We also note that the triple degeneracy of graphene-like LLs is lifted around the Lifshits transition energy $E-E_F\sim-12\meV$ (due to magnetic breakdown between different cones), so that at lower energies one can trace the fan plot of non-degenerate LLs related to a single connected dispersion surface, still well separated from the bulk states (visible in supplementary Fig. S3). At positive energies above the Dirac point we see little to no fault-localized LL formation, which happens because the stacking fault band continuously transforms into the bulk states. All these features are present in both the film calculation and in the analysis of $k_z$ bands in the "3D crystal", as compared in Fig. \ref{fig:2} (more detailed analysis of the spectra from the "3D crystal" calculation is presented in Fig.\ref{fig:sup_periodic}, including the $k_z$ band projections onto the fault layers).

\subsection{SdH oscillation frequency and cyclotron mass of the fault-related 2D band}

Having identified a distinct fault-related band, below, we analyse whether its spectroscopic and transport characteristics  would be distinguishable from the bulk states in Bernal graphite. As Shubnikov-de Haas (SdH) and de Haas-van Alphen  oscillations have been well studied in graphite \cite{Soule1964,DresselhausReview,schneider2012using,schneider10,Spry60}, we compare their established frequencies  ($6.0 <\nu_{SdH}^{e}<6.6 \T $ and $4.5 <\nu_{SdH}^{h}<4.8 \T$ for the electron and hole pockets of graphite's Fermi surface respectively) with the expected SdH oscillation frequency for the 2D band. Also, cyclotron mass in graphite  ($m_{CR} = 0.058\pm 0.001 m_0$) has been identified with the electron branch of the bulk  Fermi surface \cite{McClure1957, Galt56,Nozieres57,Inoue62,Ushio72,Suematsu72,DresselhausReview,SchneiderPotemski2009,AitorRaman}, so that we will use the 2D band cyclotron mass as another characteristic to compare 2D \textit{vs} bulk states. To mention, the SWMcC parameters used in our calculations where chosen to reproduce the above-mentioned bulk graphite characteristics (for details see supplementary section S4).

\begin{figure}
\begin{center}
\includegraphics[width=1\columnwidth]{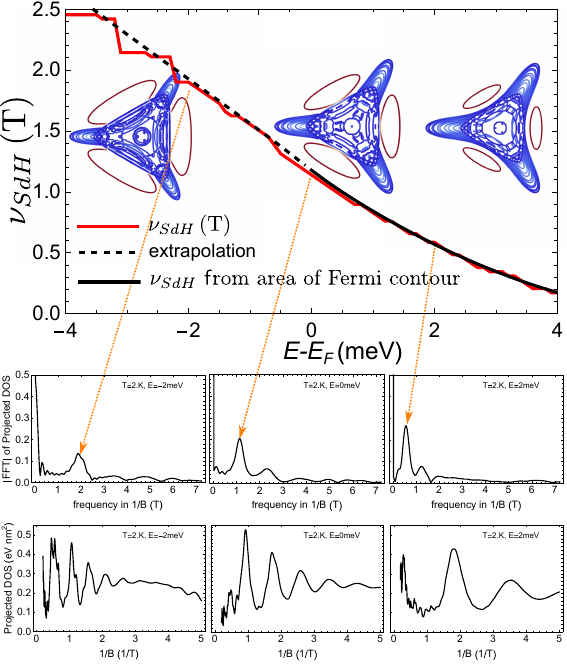}
\caption{Frequency and amplitude of SdH oscillations as a function of energy ($E=0$ corresponds to the expected Fermi level for undoped system), results were obtained by a Fourier transform in the interval $0.2 < B <5 \,\rm T$ for a 30+1+30 system. The black line corresponds to the area of the red Fermi contour when it is closed. Insets show the Fermi contours at energies $-2,\, 0,\, 2 \meV$ respectively. We observe that a SdH signal survives even in the regime when the states localized on the stacking fault do not form a closed Fermi contour: scattering to the bulk hole bands and back helps to close the path.  
\label{fig:3}}
\end{center}
\end{figure}

To analyse SdH oscillations, we make cuts of the local DoS (computed with broadened LLs) at several energies in the range $\pm 4 \meV$ near the Fermi level and plot those against $1/B$, see Fig. \ref{fig:3}.  Then, we perform a Fourier transform in the range of $1/B$ corresponding to $0.2 < B < 5 \rm T$ for each dataset, which gives us the frequency of SdH oscillations, $\nu_{SdH}(E-E_F)$, plotted in the main panel of Fig.\ref{fig:3} with the red line. The obtained frequencies agree with the areas, ${\mathcal A}(E)$ encircled by the individual Fermi contour in each of the three 2D dispersion pockets in Fig. \ref{fig:1}b and recalculated into $\nu_{SdH} = \frac{{\mathcal A}}{h e} $ (solid black line). As illustrated in the insets in Fig.\ref{fig:3}, one side of the constant-energy contour of the fault approaches bulk bands at lower energies, so that at $E<E_F$, it eventually touches the bulk bands, so that the encircled area  can only be estimated by extrapolation (dashed black line). Despite that, the frequency of SdH oscillations is clearly identifiable based on the finite-$B$ DoS data and appears to be close to the extrapolated values. Moreover, based on the Berry phase $\pi$ calculated for Dirac-like dispersion branches in Section B, we expect a $\pi$-shift of the SdH oscillations phase specific to the fault-bound bands, as compared to bulk graphite \cite{SchneiderPotemski2009}.   

\begin{figure}
\begin{center}
\includegraphics[width=1\columnwidth]{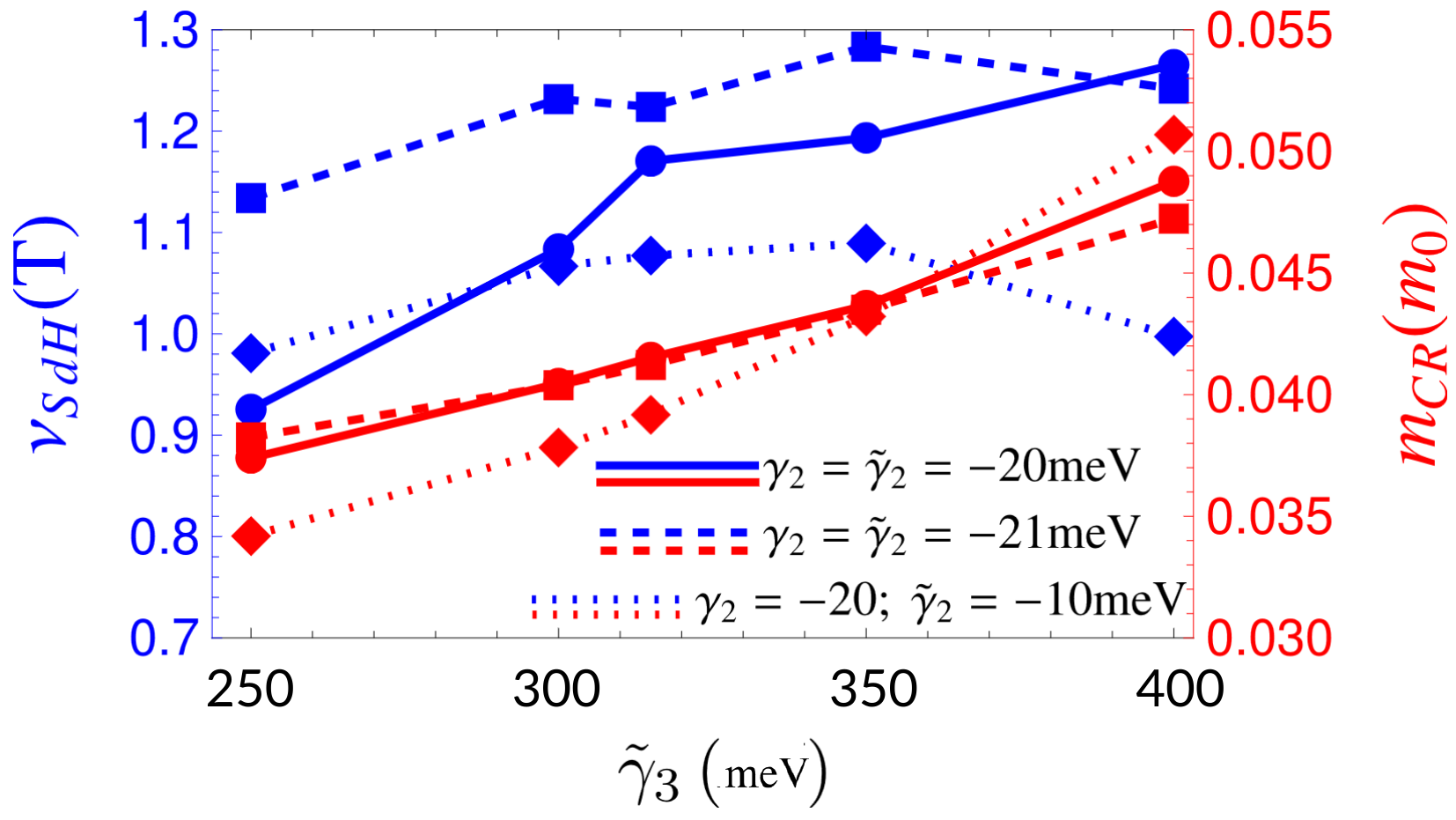}
\includegraphics[width=1\columnwidth]{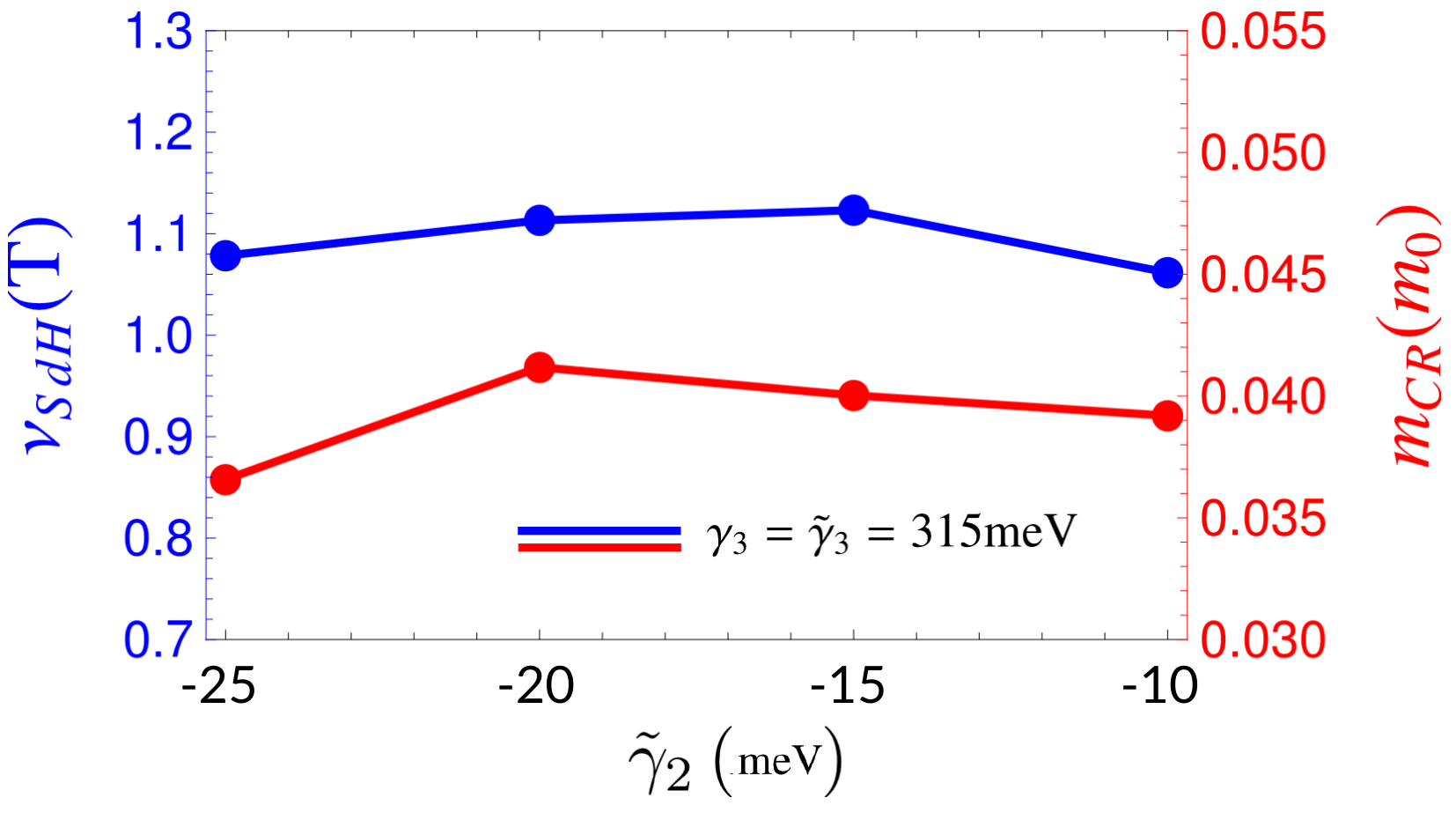}
\caption{Dependence of SdH oscillations frequency and cyclotron mass of stacking fault states on the poorly-known parameters $\tilde \gamma_3$ (a) and $\tilde \gamma_2$ (b). Blue lines are for $\nu_{SdH}$ and red lines are for  $m_{CR}$. For comparison, the bulk graphite values are $\nu_{SdH} \approx  4.5$ and $6.1 \rm T$ \cite{Soule1964,SchneiderPotemski2009} for holes and electrons, respectively,  and $m_{CR} \approx 0.058 m_0$  \cite{Potemski2012} (for electrons that dominate in CR). To mention, self-consistent on-layer potentials were calculated separately for each set of  tight-binding parameters.  
\label{fig:4}}
\end{center}
\end{figure}

\begin{figure}
\begin{center}
\includegraphics[width=1\columnwidth]{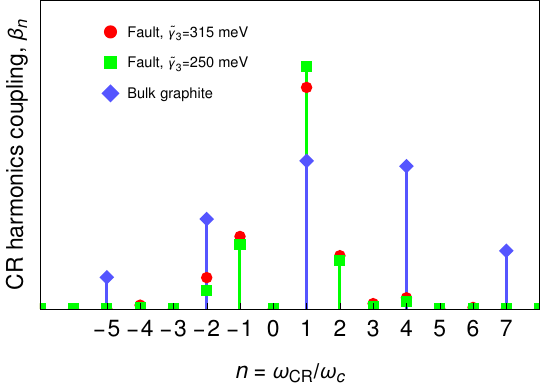}
\caption{Relative strength of CR harmonics for a stacking fault band in circularly-polarized light. Negative harmonics correspond to the opposite directions of cyclotron motion {\it vs} circular polarization of light. For comparison, a plot for graphite, extracted from experimental data in \cite{Potemski2012} is also presented.
\label{fig:CR}}
\end{center}
\end{figure}

For  the effective mass, $m_{CR}$, of 2D holes in the fault band, we use its relation \cite{AndoFowlerStern},  
\beq
m_{CR} = 2 \pi \hbar^2 \text{DoS} =  \frac{\hbar^2}{2\pi}\int\limits_{\text{pocket}} \delta(E({\bm p})-E_F)d^2{\bm p}, 
\eeq 
to  the  density of states (DoS) determined for a single 2D dispersion pocket at the Fermi level. The calculated mass agrees well  with the computed inter-LL separation, $\hbar \omega_c= \hbar e B/m_{CR}$, (for pairs of LLs closest to the Fermi level) across the magnetic field interval $0.1\T<B<1 \T$.
  For the spectrum shown in Fig.\ref{fig:2} this yields $m_{CR} = 0.042 m_0$, which is about 2/3 of the bulk cyclotron mass in graphite. 

As mentioned in Subsection
 \ref{sec:2A}, the interplane couplings $\gamma_3$ and $\gamma_2$ could be different for the layers adjacent to the fault as compared to SWMcC parameters in the bulk. For this reason we explore the sensitivity of the values $\nu_{SdH}$ and $m_{CR}$ to the choice of $\tilde\gamma_3$ and $\tilde\gamma_2$ parameters.
 In Fig.\ref{fig:4} we  illustrate those dependences, which indicate that within a reasonably wide range of $\tilde\gamma_{2,3}$ the sizes of the analysed fault observables remain distinctively different from the  values of the corresponding characteristics of bulk graphite.

We also note that inter-Landau-level selection rules for cyclotron transitions in the 2D bands differ from those of bulk graphite (where trigonal warping of dispersion permits super-harmonics $\omega_{CR}/\omega_c = \pm 1 + 3 N$, $N=...,-2,-1,0,1,2 ,...$).  To analyse the optical oscillator strength, $\beta_n$, of various cyclotron harmonics, we employ the  kinetic equation approach developed in Ref.\cite{Ushio72}, which relates coupling of 2D carriers with light characterised with a polarization vector $\bm l$ as,  
\beqa
\beta_n &\propto&  \left | \int_0^{2 \pi}  {\bm l} \cdot {\bm v}_{\bm p(s)} e^{-i n s} \frac{ds}{2 \pi}\right|^2;\\
ds &=& \frac{dp_\parallel}{m_{CR} |{\bm v}_{\bm  p(s)}|}  \ \ ,\ {\bm v}_{\bm p} \equiv {\bm \nabla_p E} .
\eeqa
Here $s/\omega_c$ is the time of flight along the cyclotron orbit in real space\cite{AbrikosovMetals} (the latter has the shape of the Fermi contour rescaled by $1/(e B)$ and rotated by $90^\circ$), which we use to parameterize the time-dependence of the instant momentum of the charge carrier at the Fermi level.   The resulting strengths of couplings with circularly-polarized light at frequencies $\omega_{CR} = n \omega_c$ are shown in Fig.\ref{fig:CR} in comparison with the relative strengths of various cyclotron harmonics detected in bulk graphite \cite{Suematsu72, Potemski2012}. This comparison shows that, independently of the fault SWMcC parameters, the CR of the holes in the 2D  band of the fault is dominated by the principal cyclotron lines $\omega_{CR} = \pm \omega_c$ with weak satellites only at $\pm 2 \omega_c$.

\section{Conclusions}
Overall, we used the tight-binding model of graphite, based on the full set of the Slonczewski-Weiss-McClure parameters, to show that a stacking fault in Bernal graphite hosts a 2D band separated from the continuum spectrum of bulk graphite, in both valleys. In particular, we found that this 2D band determines three distinct Fermi lines for the holes, which lie outside the Fermi surface of Bernal graphite. Our analysis predicts  the frequency of Shubnikov-de Haas  oscillations, $\nu_{SdH}\approx 1\T$, which substantially differs from the bulk values for graphite, $\nu_{SdH}^{e}\approx 6.3\T$ and $\nu_{SdH}^{h}\approx 4.6\T$, which are reproduced by the same tight-binding model.
We also predict the cyclotron mass of fault-bound 2D electrons to be  about 2/3 of the cyclotron mass for bulk carriers. In addition, the spectral analysis extended over a broad range of magnetic fields enables us to demonstrate the Dirac-like character of these  three separate dispersion branches at energies above the Fermi level as well as the Lifshitz transition into a single connected Fermi line at energies $\sim 15 \meV$ below $E_F$. These Dirac-like spectral features of the 2D band comply with  Berry phases $\pi$  computed for each of the three Fermi lines and the overall Berry phase $3 \pi$ calculated for a constant energy contour below the Lifshitz transition energy, which is inherited from the Berry phase properties of ABC trilayer as a determining feature of the fault structure.

\section{Acknowledgements}
We thank Leonid Glazman, Marek Potemski, and Artem Mishchenko for useful discussions. This work was supported by  EPSRC grants EP/S030719/1 and EP/V007033/1, Graphene-NOWNANO CDT, British Council Grant 1185409051 and International Science Partnerships Fund for supporting research collaboration between UK and Japan.


\appendix
\begin{center}
\textbf{\large Supplemental Materials: Title for main text}
\end{center}

\setcounter{equation}{0}
\setcounter{section}{0}
\setcounter{figure}{0}
\setcounter{table}{0}
\setcounter{page}{1}
\makeatletter
\renewcommand{\theequation}{S\arabic{equation}}
\renewcommand{\thefigure}{S\arabic{figure}}
\renewcommand{\thesection}{S\arabic{section}}

\section{Hamiltonian for graphite with ABC stacking fault}\label{Sec.SI}

\begin{figure}[hbt!]
\begin{center}
\includegraphics[width=0.8\columnwidth]{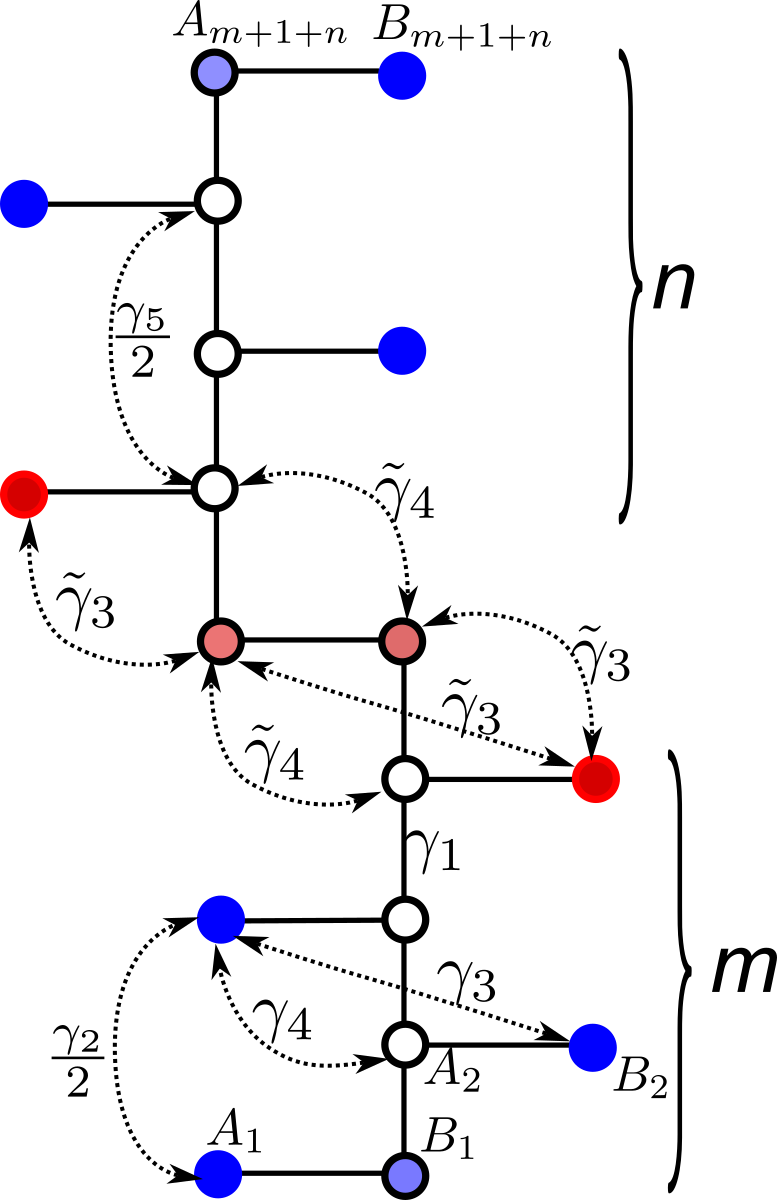}
\caption{Sketch of $m+1+n$ system with a stacking fault on layer $m+1$ (in this sketch, m=n=3). 
\label{fig:Setup}}
\end{center}
\end{figure}

We illustrate the ABC stacking fault in Bernal graphite again in Fig.\ref{fig:Setup}: there are $m$ and $n$ AB-stacked layers separated by a single layer where the string of vertical ABAB  alignment of atoms is terminated and shifted. In the terminology of Ref. \cite{KoshinoMixed13}, this is an $(m,1,n)$ stack.    

\begin{figure}[hbt!]
\begin{center}
\includegraphics[width=0.8\columnwidth]{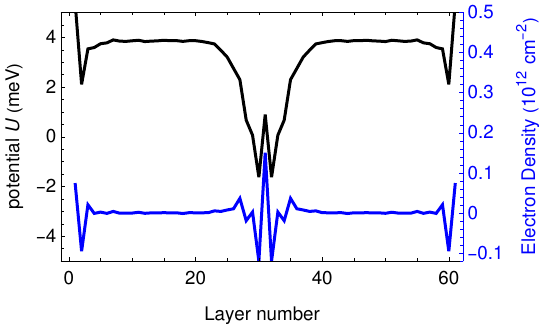}
\caption{Self-consistent Hartree potentials and on-layer charge densities computed for 30+1+30 layer system having a stacking fault in the middle. 
\label{fig:Hartree}}
\end{center}
\end{figure}

In the basis of sublattice amplitudes for electron states in  A and B sublattices of an $N=m+1+n$ film , ${\Psi}^{\dagger}=\left(
\begin{matrix}\psi_{A_1},&\psi_{B_1},&\cdots,& \psi_{A_N},
\psi_{B_{N}}\end{matrix}\right)^{\dagger}$, the $k\cdot p$ Hamiltonian is written as
\beqa \label{Hamiltonian}
\mathcal{H}&=&
\left(
\begin{matrix}
H_+^s&V^\dagger& W & 0& 0 &0& 0 &0 &0 \\
V&  H_-&V& W & 0& 0 &0& 0  &0\\
W^\dagger& V^\dagger&H_+&V^\dagger& W& 0 & 0 & 0 &0\\
0& W^\dagger& V& H_-&V_2& \tilde W& 0 &0 &0\\
0&0& W^\dagger & V_2^\dagger & H_\mathrm{g} & V_1 & W & 0 &0 \\
0& 0 & 0 & \tilde W^\dagger&  V_1^\dagger & H_+ &  V^\dagger & W &0 \\
0& 0 & 0 & 0 & W^\dagger& V& H_-&V& W \\
0 & 0& 0 & 0 & 0& W^\dagger& V^\dagger & H_+ & V^\dagger  \\
0 & 0& 0 & 0& 0 & 0& W^\dagger& V & H_-^s 

\end{matrix}
\right), \nn \\
&&H_+=H_{\mathrm{g}} + \left(\begin{matrix} 2 \Delta' &0 \\ 
                                     0 & 0 \end{matrix} \right) ,\ 
H_-=H_{\mathrm{g}} +\left(\begin{matrix} 0 &0 \\ 
                                     0 & 2 \Delta' \end{matrix} \right) 
\\&&  \nonumber
H_+^s=H_{\mathrm{g}} + \left(\begin{matrix} \Delta' &0 \\ 
                                     0 & 0 \end{matrix} \right) , \ 
H_-^s=H_{\mathrm{g}} +\left(\begin{matrix} 0 &0 \\ 
                                     0 & \Delta' \end{matrix} \right),
\\
&&H_g=v_0
\left(
\begin{matrix}
0&\pi_\xi\\
\pi^*_\xi&0
\end{matrix}
\right),\quad
V=
\left(
\begin{matrix}
-v_4\pi_\xi^*&\gamma_1\\
-v_3\pi_\xi&-v_4\pi_\xi^*
\end{matrix}
\right),
\nonumber\\
&& V_1=
\left(
\begin{matrix}
-\tilde v_{3}\pi_\xi^*&\gamma_1\\
-\tilde v_{3}\pi_\xi&-\tilde v_{4}\pi_\xi^*
\end{matrix}
\right),
V_2=
\left(
\begin{matrix}
-\tilde v_{4}\pi_\xi^*&\gamma_1\\
-\tilde v_{3}\pi_\xi&-\tilde v_{3}\pi_\xi^*
\end{matrix}
\right),
\nonumber\\
&&
\tilde W=
\left(
\begin{matrix}
0&0\\
\gamma_2/2&0
\end{matrix}
\right),\quad
W=
\left(
\begin{matrix}
\gamma_5/2&0\\
0&\gamma_2/2
\end{matrix}
\right), \nonumber \\ 
&& v_i = \frac{\sqrt{3}a}{2 \hbar} \gamma_i .\nonumber
\eeqa
where we show an explicit example for $N=3+1+3$ system shown in Fig. \ref{fig:Setup}, while the generalization to thicker Bernal stacks is done by straightforward repetition of $V^\dagger , V$ and $ H_-,  H_+$ patterns along the diagonals.  Here,  $\pi_\xi\equiv\xi p_x+ip_y$, with $\boldsymbol{p}=(p_x,p_y)$ being the valley momentum measured from $\hbar\boldsymbol{K}_\xi=\hbar\xi
\frac{4\pi}{3a}(1,0)$. 
Matrices $V$ and $W$  describe the nearest and next-nearest layer couplings, and they are assumed to be independent of the distance to the surface layers. Below we use the following values of parameters implemented in \eq{Hamiltonian}, with graphene lattice constant $a=0.246 \nm$ and  hopping energies $\gamma_0 = 3.16 \eV$, $\gamma_3 = 315 \meV $, $\gamma_4 = 44 \meV$, $\tilde\gamma_4 = 44 \meV$,  $\tilde \gamma_{3} = 315 \meV$, $\gamma_1=375 \meV$, $\Delta'=25 \meV$, $\gamma_2=-20 \meV$, $\gamma_5=38 \meV$
\cite{Gamma3Sign,yin_dimensional_2019, Ge2021Control,Potemski2012}; the hopping parameters on the stacking fault are expected to be similar to ABC stacked graphene multilayers and have not been precisely fixed in the literature we choose the same values as for Bernal graphite for our main figures and study the dependence of CR and SdH oscillations on $\tilde \gamma_{2,3}$ in Fig. \ref{fig:4}. 
The parameter $\Delta'$ accounts for energy difference of dimer and non-dimer sites (pale blue/red  vs filled  circles in Fig.\ref{fig:Setup}  and we assume that this energy difference doubles if there are Carbon atoms both directly above and below the given one (white circles in Fig.\ref{fig:Setup}), as is the case in Bernal stacking.   
In addition to SWMcC Hamiltonian, we account for the mean-field electrostatic interactions between the layers by including self-consistently calculated on-layer potentials $U_i$, $i=1,..n+m+1$ at the diagonal, accounting for the vertical electric field between the layers similarly to \cite{SlizovskiyDielectric}.  The results for self-consistent potential and charge density induced on graphene layers is shown in Fig. \ref{fig:Hartree}. There, the potential saturates to $\approx 4 \meV$ away from the defect and surfaces, corresponding to graphite Fermi level $- 4\meV$ ( or, $-4 + \gamma_2 = -24 \meV$ in the notations typical for graphite literature). The stacking fault is polarized with hole-doping on the middle layer and electrons on the adjacent layers.

The effect of an out-of-plane magnetic field $B$ is included by implementing the Luttinger substitution \cite{luttinger_quantum_1956}. We express all momentum terms in Hamiltonian (S1) by using the raising ($\hat{a}^\dagger$) and lowering ($\hat{a}$) operators of the quantum harmonic oscillator,  
\begin{equation} \notag
    \pi_{+}=\hbar\frac{\sqrt{2}}{l_B} \hat{a},
    \quad\quad
    \pi_{+}^{\dagger}=\hbar\frac{\sqrt{2}}{l_B} \hat{a}^{\dagger},
\end{equation}
\begin{equation} \notag
    \pi_{-}=-\hbar\frac{\sqrt{2}}{l_B} \hat{a}^{\dagger},
    \quad\quad
    \pi_{-}^{\dagger}=-\hbar\frac{\sqrt{2}}{l_B} \hat{a},
\end{equation}
where $l_B=\sqrt{\hbar/eB}$ is the magnetic length. To compute Landau levels, we use the basis $\{\vert n \rangle_{l}^\mathrm{A},\vert n \rangle_{l}^\mathrm{B}\}$, where $l$ is the layer index and $\ket{n}$ is the eigenstate of order $n$ of the harmonic oscillator \cite{Frank_Handbook_2010}, such that $\hat{a}^{\dag}\ket{n}=\sqrt{n+1}\ket{n+1}$ and $\hat{a}\ket{n}=\sqrt{n}\ket{n-1}$. The series is truncated at a level $N_c$ sufficiently large to ensure spectral convergence. As noted in Ref. \cite{zhang_magnetoelectric_2011}, the maximum value of $N_c$ also depends on the sublattice and layer index, this avoid artificial zero-energy LLs.

\begin{figure}[hbt!]
    \centering
    \includegraphics[width=0.8 \columnwidth]{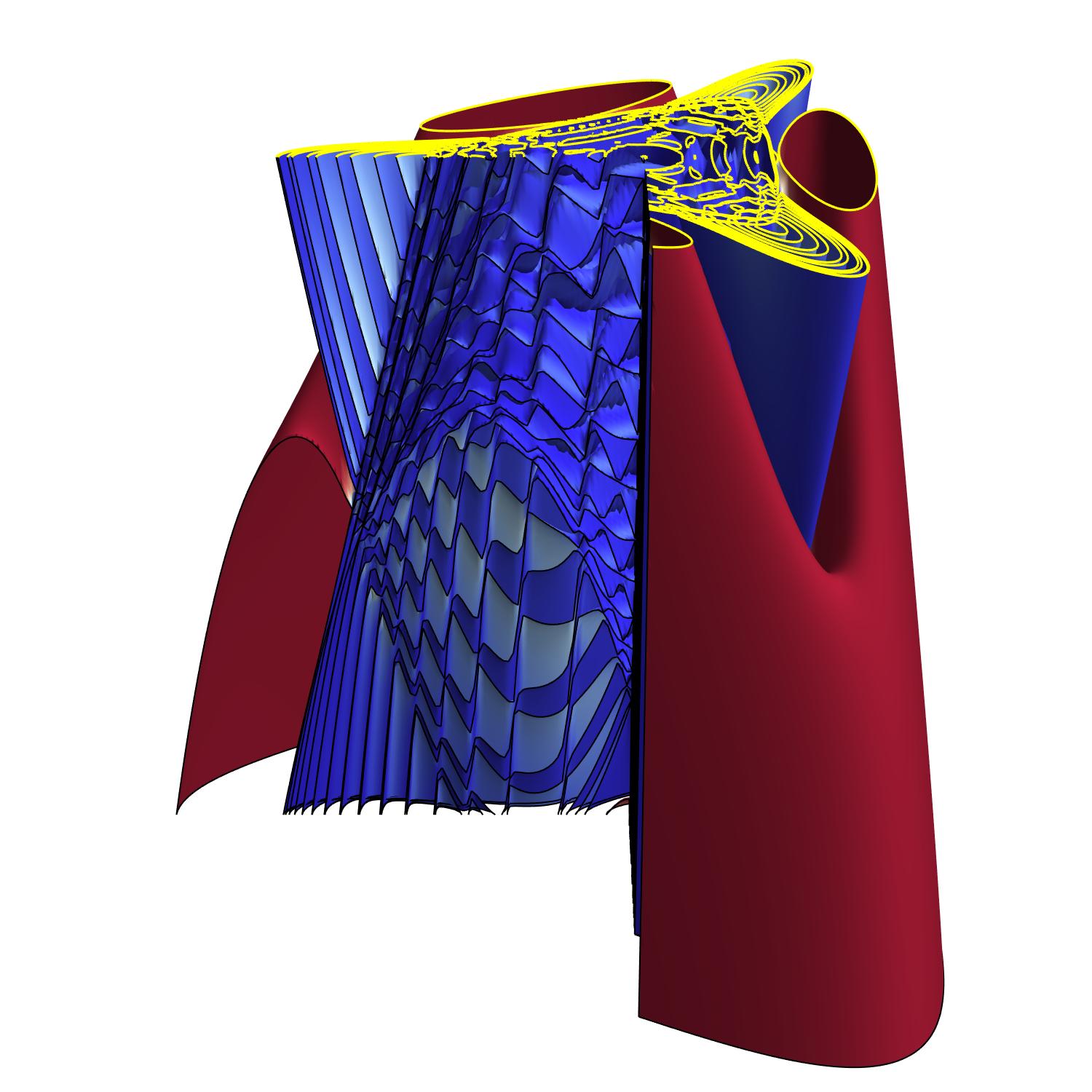}
    \caption{Result of self-consistent Hartree calculation of bands for 30+1+30 system. The red color indicates the band localized at  atomic orbitals of layers 30,31,32). The  image is cut at the Fermi energy at overall charge-neutrality. }
    \label{fig:SbandStructure}
\end{figure}

\begin{figure}[hbt!]
    \centering
    \includegraphics[width=0.8 \columnwidth]{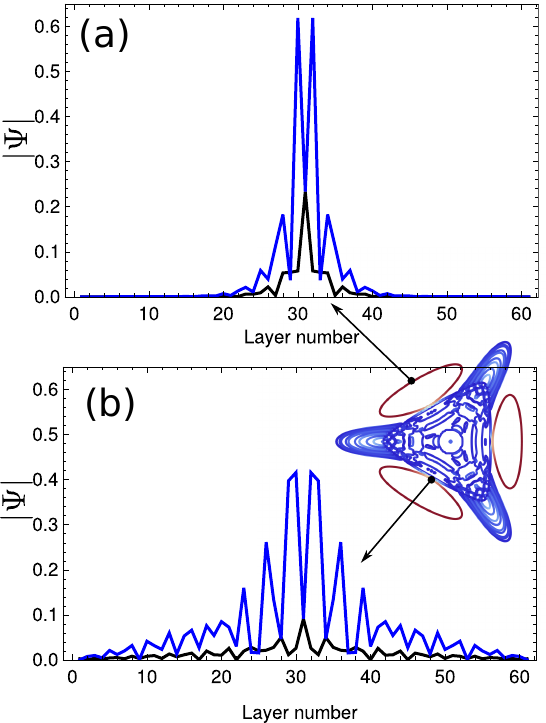}
            \caption{Amplitude of wave-function for localized state band, plotted vs the layer number. Blue/black lines correspond to wave-function amplitudes on non-dimer/dimer  sublattices. (a) is a strongly localized state far away from the bulk bands and (b) is a weaker-localized state close to bulk bands.} 
    \label{suppfig:localization}
\end{figure}

We show the result for 30+1+30 system in Fig. \ref{fig:1}, where we observe three Dirac points centered  away from the K point, in agreement with the expectations for ABC-stacked graphene trilayers\cite{Koshino2009trigonal}. The Dirac points are protected by the inversion symmetry of the stacking fault. In Fig. \ref{fig:SbandStructure}, we present the same band structure as is shown in Fig. \ref{fig:1} exposing the inner bulk bands, showing the separation of the 2D band from the bulk bands. At charge neutrality, three hole Fermi pockets are populated around these Dirac points by states localized on the stacking fault. The typical degree of c-axis localization for these state is illustrated in Fig.\ref{suppfig:localization}, where we see that probability amplitude decays by roughly a factor of 3 per layer away from the defect, but the localization weakens when the stacking fault band approaches the bulk bands of graphite.

\begin{figure}[hbt!]
    \centering
    \includegraphics[width=0.8 \columnwidth]{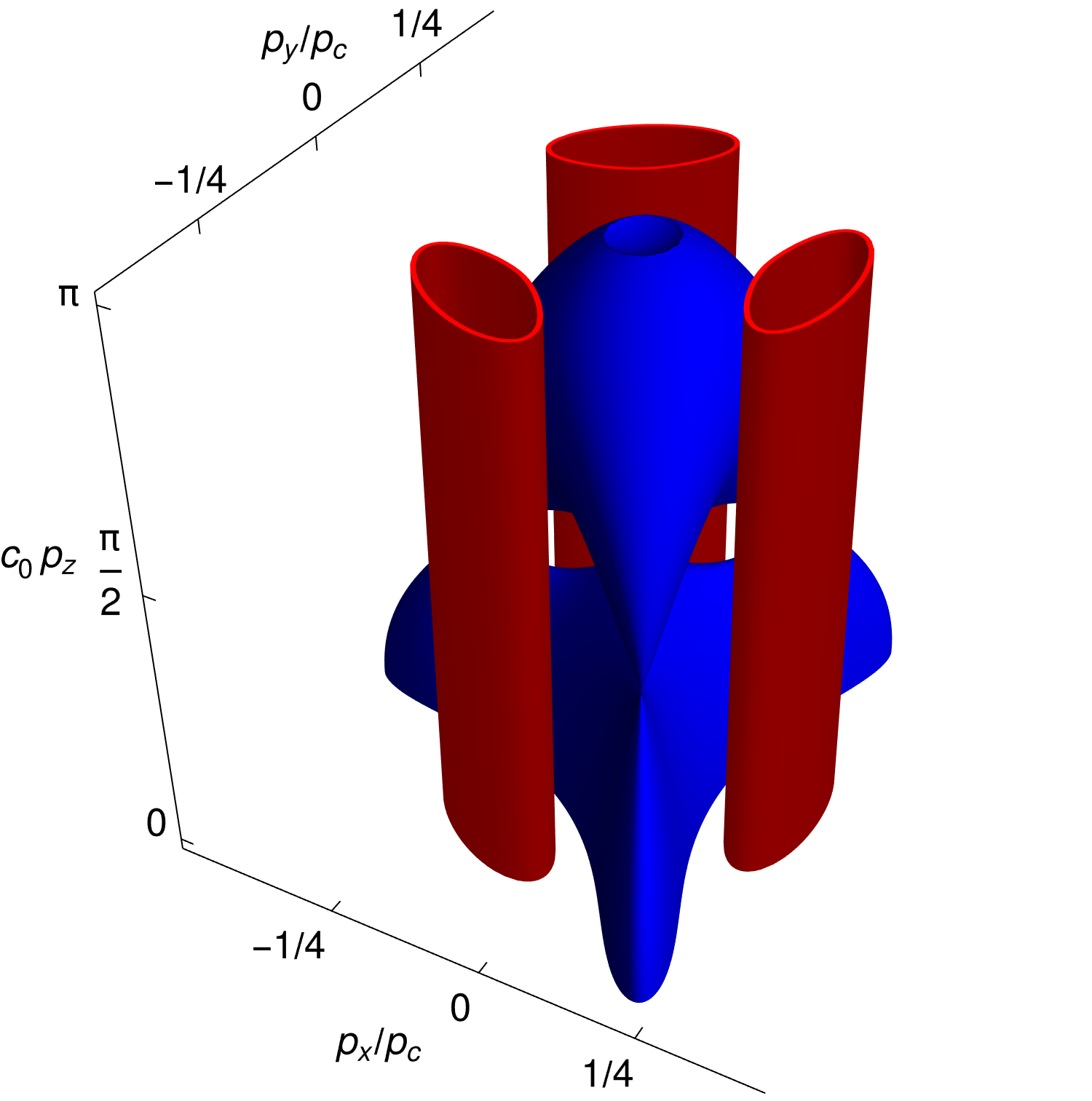}
    
        \caption{Fermi contours of 2D staking fault states (red) plotted with Fermi surface of bulk Bernal graphite (blue). Here $p_c = \gamma_1/(\hbar v_F)$. } 
    \label{fig:FS}
\end{figure}

In Fig.\ref{fig:FS} we illustrate how the Fermi surfaces of the 2-dimensional states on the stacking fault coexist with the Fermi surface of bulk Bernal graphite.

\section{Berry phase of the 2D band}
\begin{figure*}[]
\includegraphics[width=0.8\textwidth]{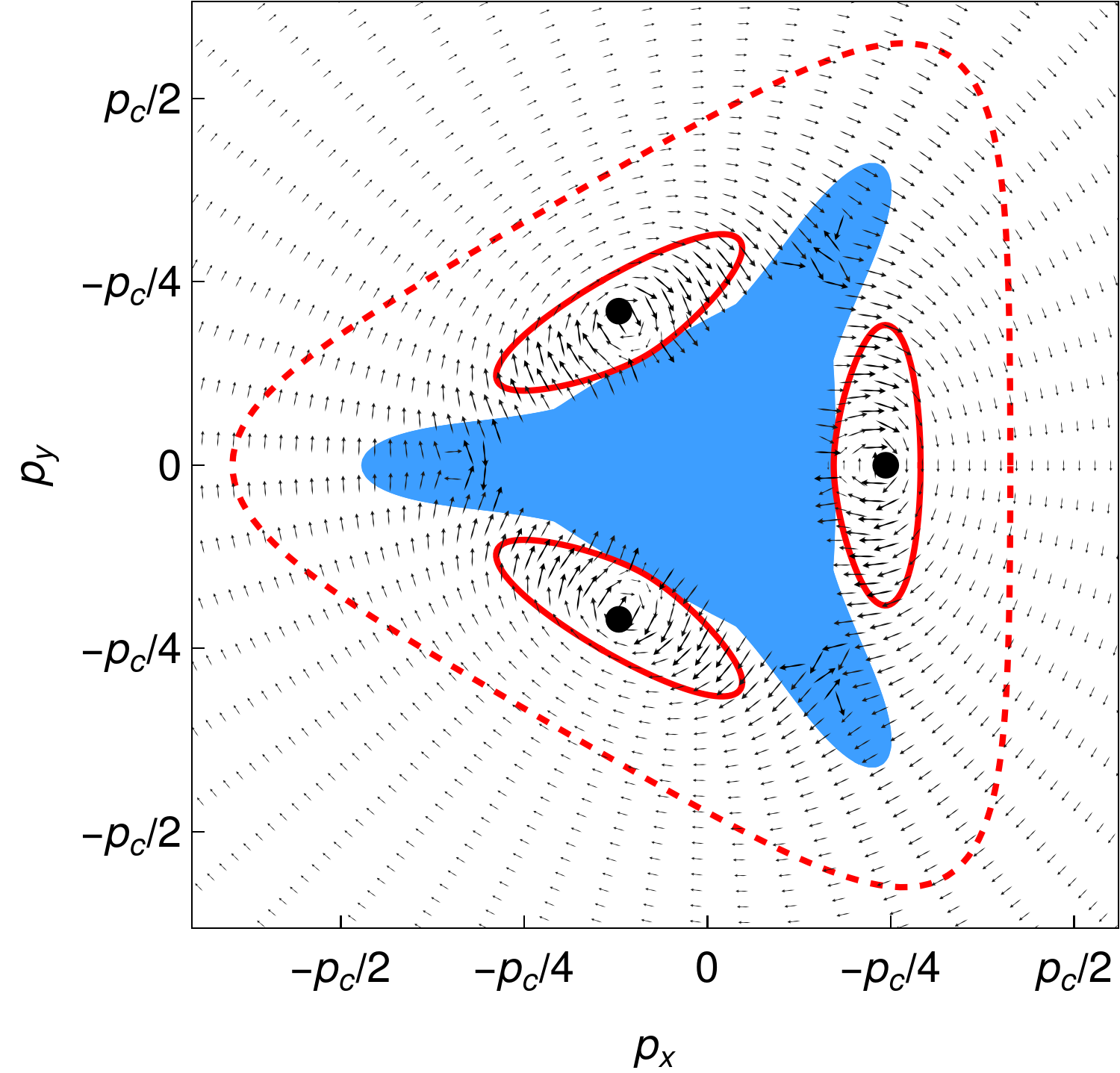}
\caption{ \label{sfig:Berry}
Vector plot of Berry connection for the fault band, $\bm A$,  along with the constant energy contours at $E=E_F$ (red solid) and $E=E_F-17 \meV$ (red dashed) and the three Dirac points (black dots). Berry connection integrates to $\phi = \mp 3 \pi$ over dashed contour  and  $\phi = \mp \pi$ over each of the smaller pockets, corresponding to the number of enclosed Dirac points. Blue area in the center corresponds to the projection of bulk Fermi surface.  
}
\end{figure*}

We calculate the Berry phase for the band localized on a stacking fault as a contour integral of Berry connection, $\bm A$,  
\begin{equation}
\label{Berry}
    \phi = \oint {\bm A} \cdot d{\bm p} \equiv -\oint \Im \langle \Psi|\nabla_{\bm p} |\Psi\rangle \cdot d{\bm p}  
\end{equation}
around a chosen fixed-energy contour.  The gauge for $\bm A$  was fixed by demanding that the wave-function on the sublattice $2 n$ (one of the red dots in Figs. 1a and \ref{fig:Setup})  is real and positive.
In Fig.\ref{sfig:Berry} we show the Berry connection, $\bm A$, together with the constant-energy contours for $E=E_F$ (red solid) and $E=E_F - 17 \meV$ (below the Lifshitz transition). We find that upon one revolution around a single Fermi pocket at $E=E_F$ the wave function acquires a Berry phase of $\pi$. In contrast, for a revolution around the large single Fermi pocket at $E=E_F-17\meV$, the wave function acquires a Berry phase of $3\pi$. 
In the opposite valley the  Berry connection is reversed. 
This result is in fact also true for the ABC trilayer (see section S4) when considering contours above and below the Lifshits transition.

\section{Dispersion and LLs for periodic array of stacking faults.}

\begin{figure*}[]
\centering
\begin{tikzpicture}

 \node (img1)
  { \includegraphics[width=1\textwidth]{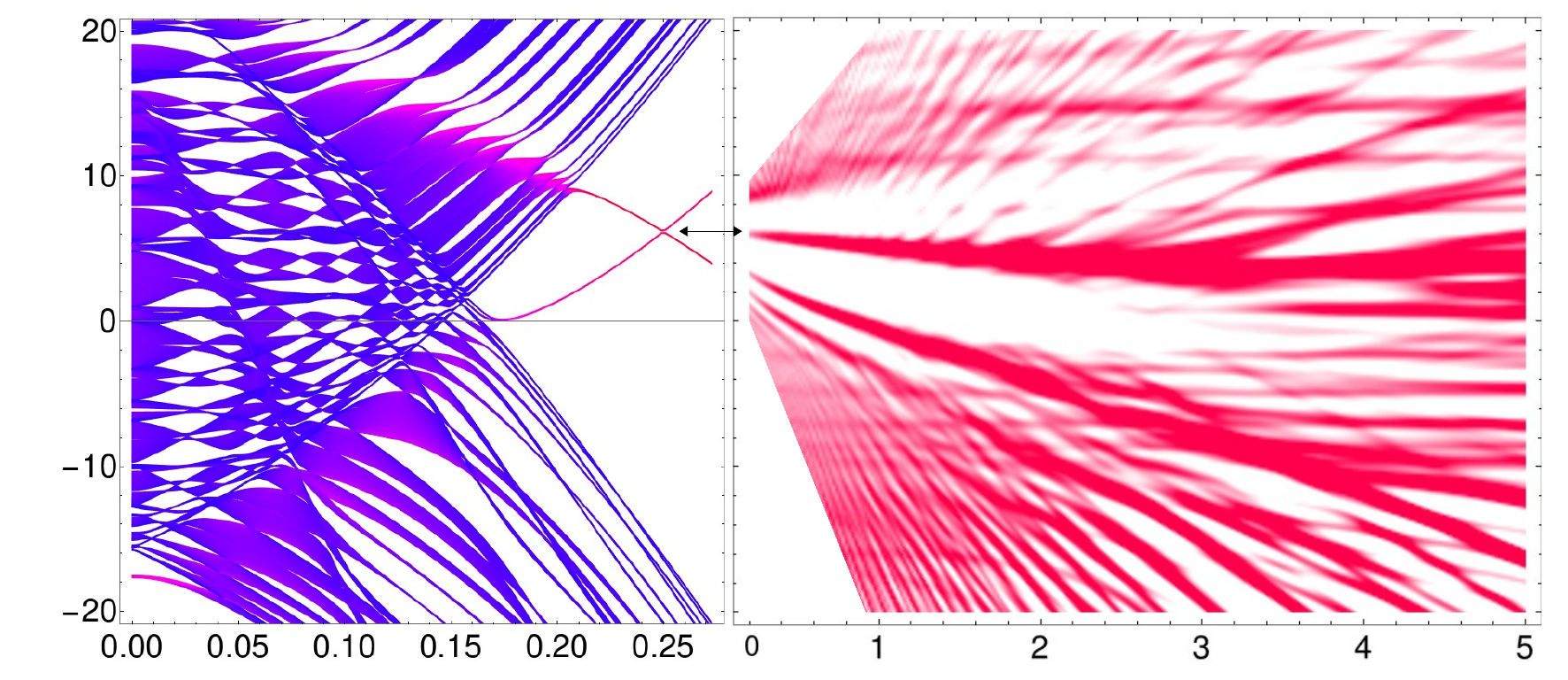}};

\node[distance=0cm,xshift=4cm, yshift=-4.1cm,font=\color{black}] {$B$ (T)};
\node[distance=0cm, rotate=90, anchor=center,xshift=0.2cm,yshift=8.5cm,font=\color{black}] {$E$ (meV)};
\node[distance=0cm,xshift=-4.0cm, yshift=-4.1cm,font=\color{black}] {$p_{x}/p_{c}$};
\end{tikzpicture}
\caption{\label{fig:sup_periodic}
(left) Bands with $p_y=0$ for an infinite periodic array of stacking faults with all values of $p_z$ superimposed, leading to finite-thickness bulk bands. Red-blue color-scale indicate projection onto ABC stacking faults / bulk respectively. We see the 2D Dirac dispersion corresponding to ABC localized states that do not disperse with $p_z$. (right) Density of states at finite magnetic field projected on to the ABC-ordered trilayer (stacking fault).   
}
\end{figure*}

The band cross-section shown in the left panel of Fig. \ref{fig:sup_periodic} was calculated using the "3D crystal" model, i.e. an infinite periodic array of stacking faults with a period of 52 layers. This super-cell include two stacking faults of opposite orientation, (ABC), (CBA),  separated by 23 layers of Bernal stacked graphene. Slicing the bands along the $p_x$ axis like this highlights the 2D nature of the isolated Dirac-like band (in red) discussed in the main text. In contrast to the 2D band, the bulk bands (in blue) have a finite thickness due to their dispersion in $p_z$. It is this dispersion in $p_z$ and the contrasting lack of dispersion in $p_z$ of the 2D bands that makes the 2D LLs clearly visible even amongst the bulk states in the DOS plot presented in the left panel of Fig. 2 in the main text.

On the right hand side panel of Fig. \ref{fig:sup_periodic} we plot the DOS in magnetic field range $0.1-5$T, projected on the fault layers for the "3D crystal" model (in contrast to the main text where we show the full DOS for the "3D crystal" model). Since this DOS is projected on the fault, the majority of the DOS contribution originates from the 2D bands. The fault bands are 2D in either realization of the system (finite or periodic model), so the DOS is almost identical to that presented for the finite model (Fig. \ref{fig:2} in the main text).

\section{Dispersion of ABC trilayer}

\begin{figure*}[]
\centering
\begin{tikzpicture}
 \node (img1)
  { \includegraphics[width=0.7\textwidth]{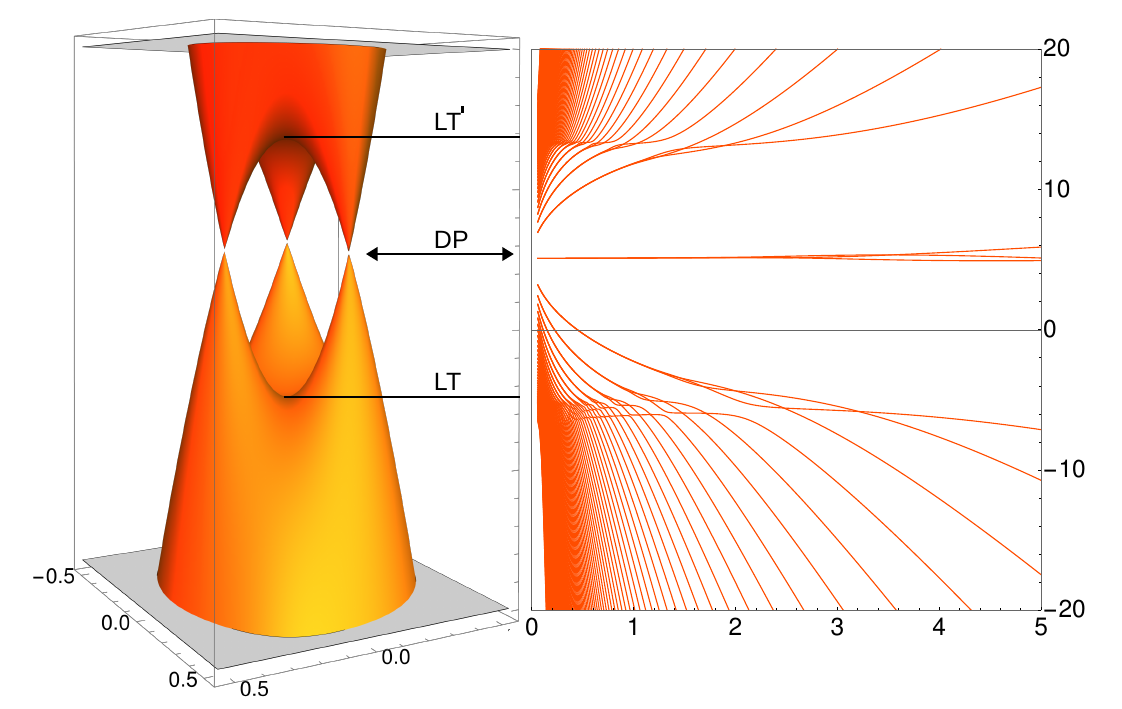}};

\node[distance=0cm,xshift=2.6cm, yshift=-3.7cm,font=\color{black}] {$B$ (T)};
\node[distance=0cm, rotate=90, anchor=center,xshift=0.2cm,yshift=-6.3cm,font=\color{black}] {$E$ (meV)};
\node[distance=0cm,xshift=-1.5cm, yshift=-3.81cm,font=\color{black}] {$p_{x}/p_{c}$};
\node[distance=0cm,xshift=-5.5cm, yshift=-3.4cm,font=\color{black}] {$p_{y}/p_{c}$};
\end{tikzpicture}
\caption{\label{fig:sup_trilayer}
(Left) Band structure of ABC graphene, calculated using the same SWMcC parameters as given in the main text. (Right) The corresponding LL spectrum. We indicate the position of the Dirac point with the black arrow which also corresponds to the same energy from which the LLs originate from. Also, the Lifshits transitions are marked by the solid black lines labeled "LT".
}
\end{figure*}

In Fig. S7, we present the band structure of ABC trilayer graphene as previously reported in \cite{Koshino2009trigonal}. One may notice the qualitative similarities between the graphene trilayer and the 2D bands from the fault layer, specifically the presence of three Dirac cones bounded by two Lifshits transitions. Correspondingly the LL spectrum which originates from these bands also displays a non-dispersive zeroth order LL accompanied by LLs with $\sqrt{B}$ dependence. All these LLs poses a triple degeneracy at the lower end of magnetic field strength which is lifted at larger fields due to the magnetic break down effect between the three Dirac cones. The Lifshits transitions are also evident in the LL spectra, much like the DOS presented in the right panel of Fig. \ref{fig:sup_periodic} for the stacking fault.

\section{Overview of bulk graphite SdH and CR results and SWMcC parameters}

In bulk graphite, SdH oscillations measure the extremal cross-sections of electron and hole parts of the Fermi surface, where the frequency of conductivity and DoS oscillations in $1/B$ is related to the extremal cross-section area $\mathcal A$ as $\nu_{SdH} = \frac{\mathcal A}{h e} $. The precise values are slightly sample- and temperature- dependent but fall in the range $6.0 <\nu_{SdH}^{e}<6.6 \T $ , $4.5 <\nu_{SdH}^{h}<4.8 \T$ \cite{Soule1964,DresselhausReview,SchneiderPotemski2009,schneider10,schneider2012using}.

The cyclotron resonance in bulk graphite corresponds to transitions between Landau levels from electron bands. The energy separation $B/m_{CR}(p_z)$ between electron-band LLs is approximately independent of $p_z$, leading to sharp peaks in light absorption. The result $m_{CR}(p_z=0) = m_{CR} = 0.058\pm 0.001 m_0$ is well established in the literature \cite{McClure1957, Galt56,Nozieres57,Inoue62,Ushio72,Suematsu72,DresselhausReview,SchneiderPotemski2009}. For hole bands, the cyclotron mass is strongly dependent on $p_z$, consequently, no sharp peaks are observed. 

\begin{figure}[hbt!]
    \centering
    \includegraphics[width=0.8 \columnwidth]{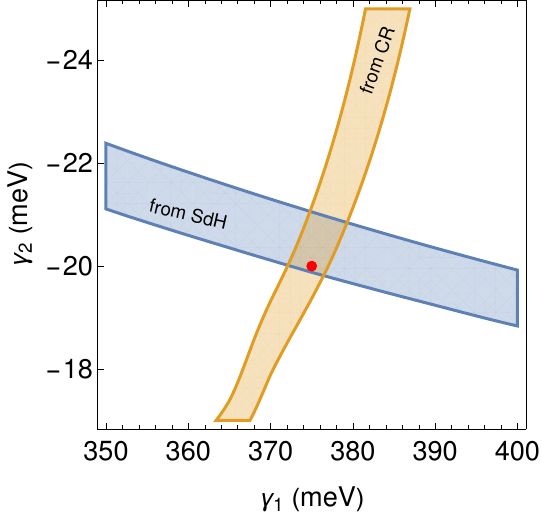}
         \caption{Restrictions on SWMcC parameters $\gamma_1$ and $\gamma_2$ coming from SdH and CR experiments. The parameters we use in the main text and those in \cite{Potemski2012} are marked with a red dot. 
    \label{suppfig:parameters} }
\end{figure}

Considering the value of $\gamma_3 = 0.315 \eV$ as fixed by previous experiments \cite{Dresselhaus2002,Gamma3Fixing} and checking that the values of $\gamma_4$ and $\gamma_5$ and $\Delta'$ are not significantly influencing our results, we are left with fine-tuning the values of $\gamma_1$ and $\gamma_2$. Since graphite samples are usually slightly doped, we have compared the sum of electron and hole SdH frequencies with theoretical predictions, as this quantity is much less sensitive to the doping. Another constraint comes from CR results. Plotting the conditions $   10.6<\nu_{SdH}^{e} + \nu_{SdH}^h < 11.3 \T $ and $0.058\, m_0<m_{CR}<0.059\, m_0$ we arrive at the result shown in Fig. \ref{suppfig:parameters}. We see that the values $\gamma_1 = 0.375 \eV, \gamma_2 = -0.020 \eV$ used in \cite{Potemski2012} are within the range and we use them in the main text.

\bibliographystyle{unsrt}
\bibliography{Bibl}

\end{document}